\documentclass[aps,prl,preprint,showpacs,superscriptaddress]{revtex4}
\usepackage{amsmath}
\usepackage{graphicx,bm}
\usepackage{color}

\newlength\figurewidth
\begin{document}

\title{Shuttle-promoted nano-mechanical current switch }
\author{Taegeun Song}
\affiliation
{Condensed Matter and Statistical Physics Section, The Abdus Salam International Center for Theoretical Physics, Trieste, Italy}

\author{Leonid Y. Gorelik}
\affiliation
{Department of Applied Physics, Chalmers University of Technology, G\"oteborg, Sweden}

\author{Robert I. Shekhter}
\affiliation
{Department of Physics, University of Gothenburg, G\"oteborg, Sweden}

\author{Mikhail N. Kiselev}
\affiliation
{Condensed Matter and Statistical Physics Section, The Abdus Salam International Center for Theoretical Physics, Trieste, Italy}

\author{Konstantin Kikoin}
\affiliation
{School of Physics and Astronomy, Tel-Aviv University, Tel-Aviv, Israel}

\date{\today}

\begin{abstract}
We investigate electron shuttling in three-terminal nanoelectromechanocal device built on a movable metallic rod oscillating between two drains. The device shows a double-well shaped electromechanical potential
tunable by a source-drain bias voltage. Four stationary regimes controllable by the bias are found for this device:
 (i) single stable fixed point, (ii) two stable fixed points, (iii) two limiting cycles, and (iv) single limiting cycle.
In the presence of perpendicular magnetic field the Lorentz force makes possible switching from one electromechanical state to another.
{\color{black} The mechanism of tunable transitions between various stable regimes based on the interplay between voltage controlled electromechanical instability and magnetically controlled switching is suggested. The switching phenomenon is implemented for achieving both a reliable \emph{active} current switch and sensoring of small variations of magnetic field.}
\end{abstract}


\maketitle

Nanoelectromechanical (NEM) systems arouse interest not only due to diverse potential applications as nano-devices but also as an efficient constituent of modern nano-electronics \cite{nemsreview1, nemsreview2}.
 While the NEM coupling plays an important part in electronic transport through nano-devices,  the charge transport associated with the nano-mechanical motion demonstrates  various interesting quantum effects, such as  Coulomb blockade \cite{cb}, resonant tunneling \cite{rt}, spin-dependent transport \cite{st} and so on.
Besides, strong NEM coupling  provides very efficient ways to control electronic and mechanical degrees of freedom of NEM-devices. {\color{black} The confined area of movable nano-meter sized island  of electron gas (quantum dot) is characterized by quantized energy spectrum.
The quantum mechanical tunneling between the source/drain and  quantum dot is responsible for a one-by-one electron charge transfer. Such electron transport by periodically moving quantum dot is known as 'shuttling phenomenon' \cite{gorelik,shuttlereview}.} The signature of shuttling was experimentally demonstrated in Refs. \onlinecite{shuttleexp1,shuttleexp2,shuttleexp3,chulki_nano}.

{\color{black} 
Recent experimental work \cite{yjunction} suggested a new type of a three-terminal NEM-device
as a current switch controlled by shifting the frequency of input signal. The "Y-switch" device consisted of three electric terminals and mechanical shuttle component - metallic island on top of nano-pillar mechanical resonator \cite{yjunction}. The three-terminal device demonstrated tunable mechanical modes operating in radio-frequency (RF) regime at room temperatures. The applications of this type of NEM-based device for quantum information processing potentially include (but not limited to) frequency dependent RF switches and ultra low-power logic elements. In our theoretical work we propose an idea of another 
three-terminal device where mechanical resonator plays also a part of one of electric terminals.
We suggest a mechanism of controlling the switching regime by magnetic field.
The high sensitivity of NEM resonator provides an opportunity to manipulate the charge transfer in  the situation when the state of device is defined by out of equilibrium conditions. These systems are referred as "active NEM devices".}

\begin{figure}
\includegraphics[width=\figurewidth]{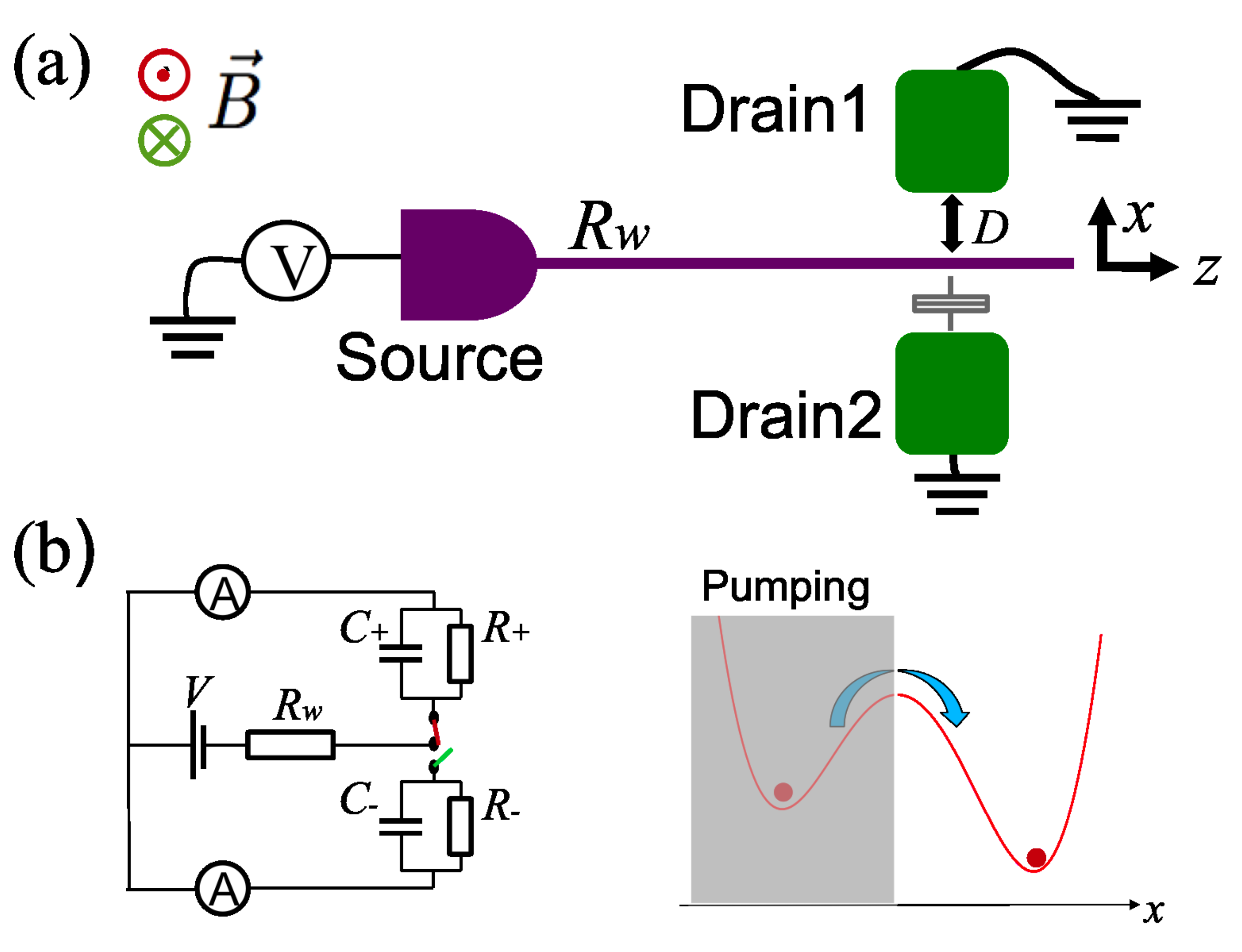}
\caption{(a) Schematic figure of the system we consider. (b) Equivalent electric circuit (left panel) and electromechanical potential $U_{\text{eff}}(x)$ in the presence of perpendicular magnetic field (right panel). (See the discussion of switching induced by magnetic field in the text).}
 \label{fig:f1}
 \end{figure}

In this Letter, we consider a NEM system containing a cantilever as a source located at the symmetric point between two vertical drains separated by the air gap of width $2D$ as shown in Fig. \ref{fig:f1} (a).
We find that dynamical behaviour of the system shows four distinguishable regimes {\color{black} of mechanical vibration} as a function of source-drain bias voltage: (i) single stable fixed point, (ii) two stable fixed points, (iii) two limiting cycles, and (iv) single limiting cycle.
Focusing on the regimes (ii) and (iii), we find conditions for the transition between two stable states tunable by switching the pulse-shaped magnetic field.

We start with formalism of equivalent circuit model for tunnel junction
combined with Newtonian dynamics of the cantilever of length $L$ vibrating within the air gap width $2D$ as shown in Fig. \ref{fig:f1} (b).
{\color{black} First, we describe mechanical degrees of freedom of the cantilever by the displacement
$u(z)$ characterizing the cantilever deflection from the straight configuration at the point $z$
(that is, the cantilever axis with its origin at the fixed end, see Fig 1.a). Second, we introduce the eigenmode representation for the cantilever displacement \cite{Santa11} and characterize the
{\it fundamental mode} of the vibration by the amplitude $x$. While two position-dependent inverse capacitances  $C^{-1}_\pm (x)$ of the parallel-plate capacitors are given by a linear function of $x$, the tunnel resistances of the air gap $R_\pm(x)$ exponentially depend on the width of the barrier: }
$$C_{\pm}(x)=C_0\frac{D}{D\mp x}, \;\;\;\;\;\;\;R_{\pm}(x)=R_0 e^{\mp x/\lambda}.$$
Here, $C_0$ ($R_0$) is capacitance (resistance) of air gap of the width $D$, and $\lambda$ is a phenomenological tunneling length. 
The force acting on the cantilever is the vector sum of electrostatic force applied to the end of the cantilever, and Lorentz force induced by magnetic field $\vec{B}=B_0\vec{e}_y$.
The current {\color{black} $I$} through the cantilever and the induced bias across the junction $V_{c}$  satisfies the Ohm's law: $I=(V-V_c)/R_w$. 
We use following notations:  
$R_w$ is resistance of the cantilever, $V_{c}=Q_c/C(x)$, $Q_c$  is total charge accumulated  inside the parallel-plate capacitors and $C(x)=C_{+}(x)+C_{-}(x)$.
Time evolution of the charge accumulated  inside the parallel-plate capacitors can be written as:
\begin{eqnarray}
\dot{Q_c}+\left(\frac{1}{\tau_c(x)}+\frac{1}{\tau_w(x)}\right)Q_c-\frac{V}{R_w}=0,
\label{chargedyn}
\end{eqnarray}
where $\tau_c(x)=R(x)C(x)$ is a position-dependent $RC$-time of the tunnel junction, $\tau_w=R_wC(x)$.
Here $R({\color{black} x})=[1/R_{+}(x)+1/R_{-}(x)]^{-1}=R_0(2\cosh(x/\lambda))^{-1}$.
Then, the electrostatic force applied to the cantilever is given by 
$\vec{F}_c=-(Q_c)^{2}(\partial C^{-1}(x)/\partial x)/2 \cdot \vec{e}_x$, and
the effective Lorentz force induced by the the current is $\vec{F}_{b}=L \vec{I}\times \vec{B}$.
Thus, the equation of motion for the vibrating cantilever within the air gap along $\hat{x}$-direction is given by,
{\color{black}
\begin{eqnarray}
\ddot{x}+2\gamma_0\dot{x}+\omega_0^2 x=a_1\frac{Q_c^2}{C_0 D^2 m} x+a_2\frac{LB_0}{m R_w}\left(V-V_c\right),
\label{equmot1}
\end{eqnarray}
}
where $m$ is the effective mass, $\omega_0$ is a frequency of fundamental mechanical mode and $\gamma_0$ is 
{\color{black} its damping coefficient. Here, $a_1,a_2 \sim 1$ are geometrical factors.}

In order to present a system of coupled equations describing both mechanical motion and charge distribution
in compact form, we introduce dimensionless variables, denoted by tilde, which are defined by re-scaling {\color{black} the displacement with $\lambda$, the time with $\omega_0^{-1}$,} the current with $e\omega_0$, the voltage with $e/C_0$, and the force with $m\omega_0^2\lambda$ {\color{black} ($\tilde{x}=x/\lambda$, $\tilde{t}=\omega_0 t$, {\color{black} $\tilde q_c=Q_c/e$}, $\tilde{I}=I/e\omega_0$, $\tilde{v}=C_0 V/e$, and $\tilde{F}=F/m\omega_0^2\lambda$)}:
{\color{black}
\begin{eqnarray}
\ddot{\tilde{x}}&+&\frac{1}{\text{Q}_0}\dot{\tilde{x}}+\tilde{x}=\frac{\alpha}{d^2} \tilde{q}_c^2\tilde{x}+\frac{\pi\beta\phi_B}{\tau_0 r_w}\left(\tilde{v}-\tilde{q}_c \right),\label{xtil}\\
\dot{\tilde{q}}_c&+&\frac{1}{\tau_0 r_w}\left(r_w\cosh(\tilde{x})+1\right)\tilde{q}_c=\frac{\tilde{v}}{\tau_0 r_w}. 
\label{qtil} 
\end{eqnarray}
}
\noindent
with $\tau_0=\omega_0 R_0 C_0$, $d=D/\lambda$, $r_w=R_w/R_0$, {\color{black} $1/Q_0=2\gamma_0/\omega_0$}. {\color{black} Here} $\phi_B=\lambda L B_0/(h/e)$
{\color{black} is dimensionless flux through the area of triangle with linear sizes
determined by the length of the cantilever and amplitude of mechanical vibration measured in the units of flux quantum $\phi_0=h/e$.}
The {\color{black} dimensionless} parameters $\alpha$, and $\beta$ correspond to the charging energy $E_c=e^2/C_0$, 
and oscillator {\color{black} (zero point motion)} energy $E_{\rm osc}=\hbar\omega_0$ 
{\color{black} measured in units of elastic energy $E_{e}=m\omega_0^2\lambda^2$}:
{\color{black} $\alpha=a_1 E_c/(m\omega_0^2\lambda^2)$, $\beta=a_2 E_{\rm osc}/(m\omega_0^2\lambda^2)$.}
{\color{black} Note, that dimensionless RC-time $\tau_0$ appears in Eq.\ref{xtil} due to rescaling of voltage with the charging energy which is a "natural" unit for rescaling in Eq.\ref{qtil}.}
We ignore the bending effects of the cantilever inside the {\color{black} area between source and drain(s)}, and, {\color{black} assuming that the condition $\tilde{x}/d \ll 1$ is satisfied, disregard the corrections of the order of $(\tilde{x}/d)^2$ in the equations {\color{black} Eq.(\ref{xtil}) and Eq.(\ref{qtil})}.}

Two terms in the r.h.s. of Eq.(\ref{xtil}) account for the retardation effects due to redistribution of charge and the Lorentz force acting on the moving cantilever.
In the adiabatic limit $\tau_0\ll1$ we assume that the dynamics of the charge distribution is determined by {\color{black} $RC$- time which is much faster compared to the time scales responsible for}  dynamics of mechanical degrees of freedom. The approximate analytic solution of Eq. (\ref{qtil}) describes the position-dependent charge distribution, $\tilde{q}_c({\color{black} \tilde{x}})$:
\begin{eqnarray}
\tilde{q}_c({\color{black}\tilde{x}})=\frac{\tilde{v}}{r_w\cosh(\tilde{x})+1}. \label{qdist}
\end{eqnarray}
As one can see from Eq.(\ref{qdist}), the charge accumulated at the tip of 
the cantilever decreases exponentially with the amplitude $\tilde{x}$. 

\begin{figure}
\includegraphics[width=\figurewidth]{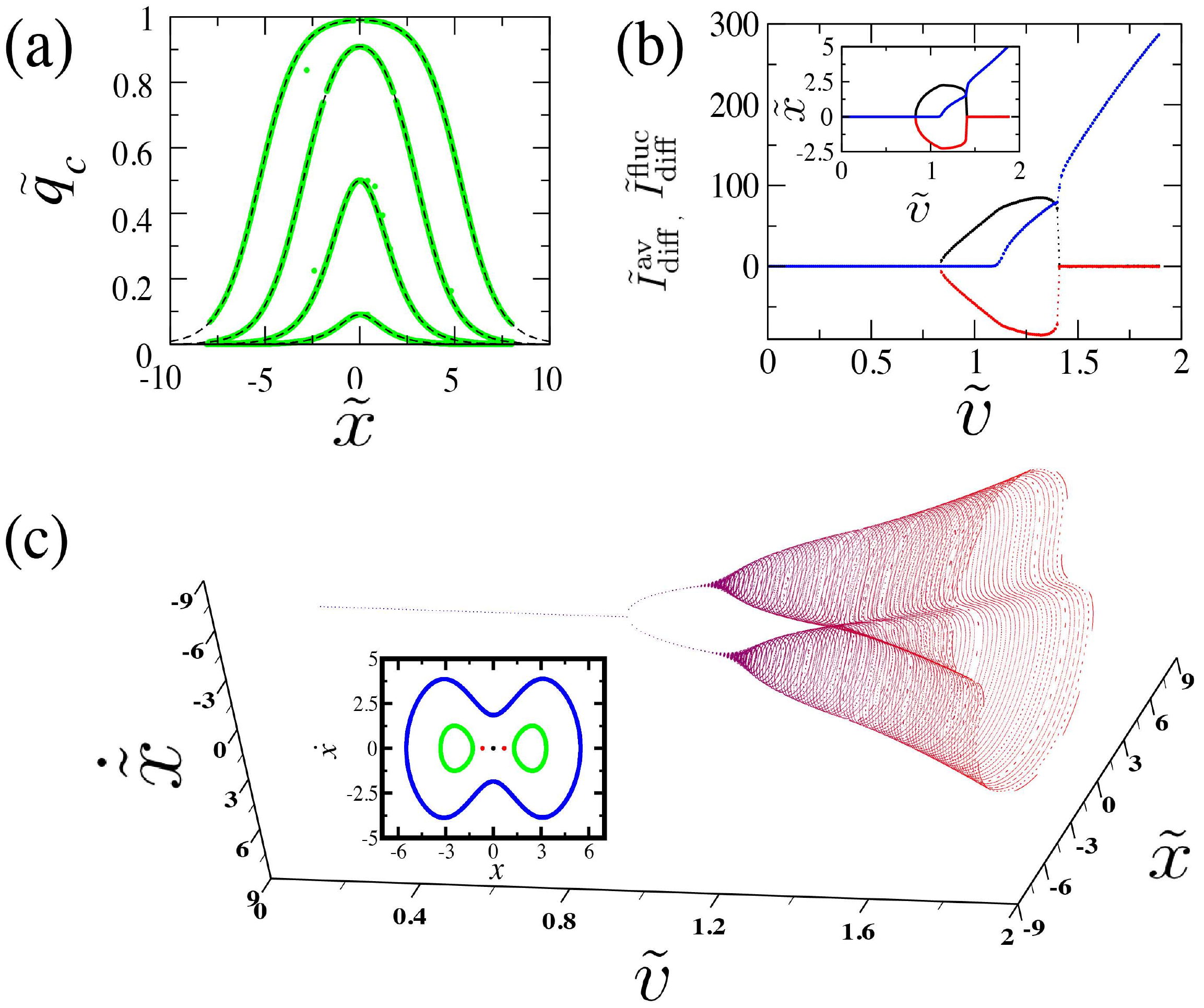}
\caption{
(a) Charge distribution $\tilde q_c$ given by Eq. 5 (black dashed lines) and 
Eq.(\ref{xtil}) - Eq.(\ref{qtil}) (green dots). The parameter $r_w=1,~0.1,~0.01,~\text{and},~0.001$ (from top to bottom).  Other parameters are fixed at the values $\tilde v=1$ and $\alpha/d^2=1.75$.
{\color{black} (b) The re-switching current ${\color{black} \tilde{I}}_{\text{diff}}^{\text{av}}$ (black and red lines) defined as  the current difference between source-drain1 and source-drain2 and its fluctuations 
$\tilde{I}_{\text{diff}}^{\text{fluc}}$ (blue line) averaged over the time interval 
$T=1000 \omega_0^{-1}$ (two different colors for re-switching current correspond to  two different initial conditions: black - oscillation near drain1 $\tilde{x}_{int}=0.1$; red - oscillations near drain2, $\tilde{x}_{int}=-0.1$) as a function of bias. 
Inset: averaged displacement relative to symmetric position of the cantilever and its fluctuations.}
(c) {\color{black} Poincar\'{e} map} of $(\tilde{x},\dot{\tilde{x}})$ {\color{black} for the steady state at zero magnetic field $B=0$  evaluated after delay time $(t>5000~\omega_0^{-1})$ as a function of bias voltage.} 
Inset: cross section of main plot at the bias voltages $\tilde{v}=0.8$ (black dot), $0.85$ (two red dots) $1.2$ (two green curves) $1.6$ (blue curve). 
The parameters are: $\tau_0=0.1,~\beta=0.01$, $\text{Q}_0=100$. 
In order to calculate (a) and (c), we choose 150 random initial conditions in the range of 
$(\tilde{x}_{int},\dot{\tilde{x}}_{int},\tilde{q}_{c\_int})\in[-5,5]$.
}
\label{fig:f2}
\end{figure}

First, let us consider the setup in the absence of perpendicular magnetic fields, $B=0$
(see also \cite{isaak}).
Then, the effective electromechanical potential, {\color{black} assuming $r_w\ll 1$}, can be written as
\begin{eqnarray}
U_{\text{eff}}(\tilde{x})=\frac{1}{2}\tilde{x}^2+\frac{\alpha}{d^2}\left(\ln[\cosh(\tilde{x})]-\tilde{x}\tanh(\tilde{x})\right).
\end{eqnarray}
The values of two local minima $\tilde{x}_{\pm}$ are found by solving equation $\partial_xU_{\text{eff}}(\tilde{x})=0$. The solution reads: $\tilde{x}_{\pm}=\pm \text{cosh}^{-1}((\tilde{v}\sqrt{\alpha/d^2}-1)/r_w)$.
The emergence of $\tilde{x}_{\pm}$ gives us the condition for threshold bias ($\tilde{v}_1$) necessary for formation of a double-well shaped $U_{\text{eff}}$ which is written as: $\tilde{v}_1 > (r_w+1)/\sqrt{\alpha/d^2}$.

Nonadiabatic correction to the adiabatic charge 
{\color{black}
\begin{equation}
\tilde{q}^{na}_c=\frac{\tilde v r_w^2 \sinh({\tilde{x}})}{(r_w \cosh(\tilde{x})+1)^3}\cdot \dot{\tilde{x}}\cdot \tau_0
\end{equation}
generates effective "negative" friction in the vicinity of the minima of the double-well potential.
 As a result further increase of bias gives rise to instability of the static state. 
Finite energy pumping generates limiting cycle at the vicinity of two local minima
depending on the initial condition.
With growing bias voltage, the two limiting cycles evolve into one large limiting cycle enveloping two local minima. This happens  when the energy pumping allows the system to overcome the barrier between two local wells.
Since there are two characteristic voltages controlling the number of limiting cycles of the system,
it is convenient to introduce two other critical voltages, $\tilde{v}_2$ and $\tilde{v}_3$ for two limiting cycles and one limiting cycle, respectively.
General expressions for $\tilde{v}_2$, and $\tilde{v}_3$ are unavailable, however, we evaluate the characteristic voltages in Supplementary materials.

{\color{black} The main focus of this paper is to demonstrate the re-switching behaviour of active device in the situation when the current injected mainly in the drain1 is eventually forced to be injected to the drain2. This is why we will be interested in the calculation of the difference between the current injected  from the source to each of the drains, calling the current through the drain1 as $I_+$ and the drain2 as $I_-$ (Fig 1.a). The difference between these two currents, which we call a "re-switching current" $\tilde I_{\text{diff}}$, can be fine-tuned by applying external magnetic field. The current ${\color{black} \tilde{I}}_{\text{diff}}({\color{black}\tilde{t}})={\color{black} \tilde{I}}_+({\color{black}\tilde{t}})-{\color{black} \tilde{I}}_-({\color{black}\tilde{t}})$ fully describes switching properties of the active device. It is convenient to characterize switching by the current averaged over a time interval $T$ being large compared to the period of mechanical vibrations (in the numerical calculations we use $T=1000~\omega_0^{-1}$). Another important for switching dynamics time scale is associated with the delay $t$ after which we perform the time averaging $(t>5000~\omega_0^{-1})$ in the steady state (see Fig. \ref{fig:f2}(b)).}
For $\tilde{v}_1\leq\tilde{v}\leq \tilde{v}_3$, the difference ${\color{black} \tilde{I}}_{\text{diff}}^{\text{av}}$ splits following the evolution of the system and depending on the initial condition  either to the right ($\tilde{x}_{int}>0$, black color in Fig. \ref{fig:f2}(b)) or to the left ($\tilde{x}_{int}<0$, red color) near one of two minima of the potential $U_{\text{eff}}(\tilde x)$.
The shuttling regime ($\tilde{v}_2\leq\tilde{v}$), is characterized by non-zero fluctuations of the current difference, ${\color{black} \tilde{I}}_{\text{diff}}^{\text{fluc}}=\langle {\color{black} \tilde{I}}_{\text{diff}}({\color{black}\tilde{t}})-{\color{black} \tilde{I}}_{\text{diff}}^{\text{av}} \rangle$.
The fluctuation strength continues to increase after a sudden drop of ${\color{black} \tilde{I}}_{\text{diff}}^{\text{av}}$ at ${\color{black}\tilde{v}}_3$, (see the blue line in Fig. \ref{fig:f2}(b)).
The stationary {\color{black} Poincar\'{e} map} of various random initial condition for $(\tilde{x},\dot{\tilde{x}})$ as a function of $\tilde{v}$ is shown in Fig. \ref{fig:f2}(c).
{\color{black} The one-to-one correspondence between the re-switching current and displacement shown on Fig. \ref{fig:f2} (b) can be used for position detection of the nano-device.}

Next, we describe the setup in the presence of perpendicular magnetic field applied in order to manipulate the switching current between source and one of two drains, ${\color{black} \tilde{I}}_{\pm}$ in the $\tilde{v}_1\leq\tilde{v}\leq\tilde{v}_3$ regime. 
We consider adiabatically 
adiabatically varying time-dependent flux
$\phi_B({\color{black}\tilde{t}})=\frac{b_0}{2}(\tanh(\frac{{\color{black}\tilde{t}}-{\color{black}\tilde{t}}_s}{\tau_p}){\color{black}-}\tanh(\frac{{\color{black}\tilde{t}}-{\color{black}\tilde{t}}_e}{\tau_p}))$, under {\color{black}following} condition {\color{black} for duration of the flux pulse $\tau_d={\color{black}\tilde{t}}_{\color{black}e}-{\color{black}\tilde{t}}_{\color{black}s}$ and saturation time $\tau_p$} {\color{black} compared to the $RC$-time $\tau_0$}: 
$1/{\color{black} \text{Q}}_0<\tau_0\ll\tau_d\ll\tau_p$.
 Fig. \ref{fig:f3} (a) to (d) illustrates switching dynamics of a system initially located in the left minimum.
We apply pulses $\tau_d=200$, and $\tau_p=1000$ at ${\color{black}\tilde{t}}_s=15000~({\color{black}\tilde{t}}_{\color{black}e}=35000)$ for stimulating jumps from the left well to the right one and back. 
{\color{black}The voltage dependence of the lower/upper critical fields is shown on Fig. {\ref{fig:f3}} (e). 
If the magnetic field exceeds its upper critical limit, the double-well potential transforms into the single-well potential (Fig. \ref{fig:f4} (b)).
{\color{black} The voltage dependence of the upper critical flux $\phi_B^{max}$ can be obtained 
by evaluating the minima of confining potential
under condition that two stable minima transfer into single stable minimum.}
The lower critical field $\phi_B^{min}$ have been numerically defined as the minimal value of applied flux by comparing $\tilde{x}$ averaged over long time scale $T$ before and after stimulation.}
{\color{black} In Fig. {\ref{fig:f3}} (e),} red colored line shows log-scaled current square average, $\langle| {\color{black} \tilde{I}}_+({\color{black}\tilde{t}})+{\color{black} \tilde{I}}_-({\color{black}\tilde{t}})|^2\rangle$ at the $\phi_B^{min}$ which is directly proportional to current power. {\color{black} It is therefore demonstrated that there exist regimes when small magnetic field can switch large currents in the active regime of nano-device due to amplification of device sensitivity by the preceding signal.}
As is seen from Fig. \ref{fig:f3} (e), small flux switches between two different regimes both at the voltages around $\tilde{v}_1$ and $\tilde{v}_3$. This means, that in addition to use of magnetic field for manipulating the current switch, one can use switching itself for detection of small magnetic fields thus providing a highly  sensitive magnetic field sensor.

\begin{figure}
\includegraphics[width=\figurewidth]{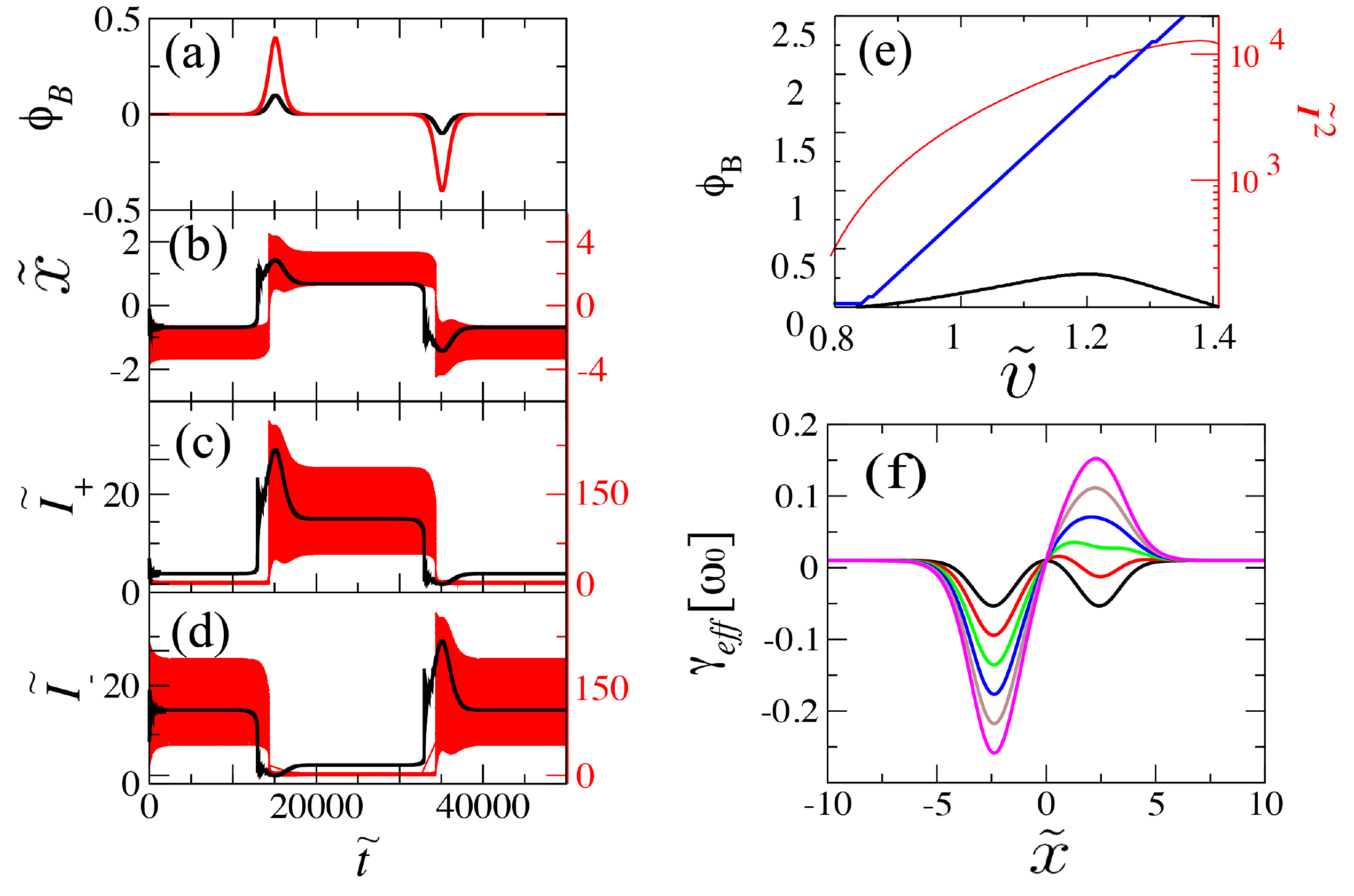}
\caption{Switching behavior of the device controlled by magnetic pulses. Time scanning of $\phi_B$ (a), $\tilde{x}$ (b), ${\color{black} \tilde{I}}_+$ (c), and ${\color{black} \tilde{I}}_-$ (d) at the bias voltage $\tilde{v}=0.85,~b_0=1$ (black), and $\tilde{v}=1.2,~b_0=4$ (red).
(e) Critical magnetic field {\color{black} $\phi_B^{min}$ (black) and $\phi_B^{max}$ (blue)}  for switching phenomenon as a function of bias and the mean-square of the total current averaged at the critical value of magnetic field (red). (f) Position-dependent effective friction under different {\color{black} values of dimensionless flux $\phi_B$ from its minimal valued $0$ (black) to maximum value $5$ (magenta) with the step $\Delta \phi_B=1$.} 
}
 \label{fig:f3}
 \end{figure}

The switching mechanism based on magnetic fields in the device can be considered by using position-dependent effective dissipation coefficient.
Since the device shows position-dependent charge distribution, the Lorentz force involves 
non-adiabatic corrections to charge dynamics.
Consequently, position-dependent effective dissipation has been emerged in the equation of motion:
{\color{black}
\begin{eqnarray}
\gamma_{\text{eff}}(\tilde{x})=\gamma_0
+\frac{\pi \beta \omega_0\tilde v \phi_B r_w \sinh(\tilde x)}
{(r_w \cosh(\tilde x)+1)^3}
-\frac{2\alpha\omega_0\tau_0}{d^2}\frac{\tilde v^2 r^2_w \tilde x \sinh(\tilde x)}
{(r_w \cosh(\tilde x)+1)^4}
\label{gameff}
\end{eqnarray}
}
Fig. \ref{fig:f3} (f) shows position-dependent $\gamma_{\text{eff}}(\tilde{x})$ as a function of varying magnetic field. Unlike the Lorentz force, which is an odd function of the coordinate $\tilde{x}$, {\color{black} flux $\phi_B$ and voltage $\tilde{v}$}, the {\color{black} non-adiabatic contribution to the electrostatic force being even function of both coordinate $\tilde{x}$ and bias voltage $\tilde{v}$} always reduces the dissipation near stationary position (see last term in Eq.\ref{gameff}).

Using adiabatic approximation, we  calculate the phase diagrams of bi-stability regime (see Fig. \ref{fig:f4}), from which the potentialities of current switch can be seen.
We use folowing color scheme in
Fig. \ref{fig:f4}(a): gray color is used for \emph{'passive'} switching regime $(\tilde{v}_1<\tilde{v}<\tilde{v}_2)$, brown color denotes the \emph{'active'} switching regime based on the shuttling instability $(\tilde{v}_2<\tilde{v}<\tilde{v}_3)$.

\begin{figure}
\includegraphics[width=\figurewidth]{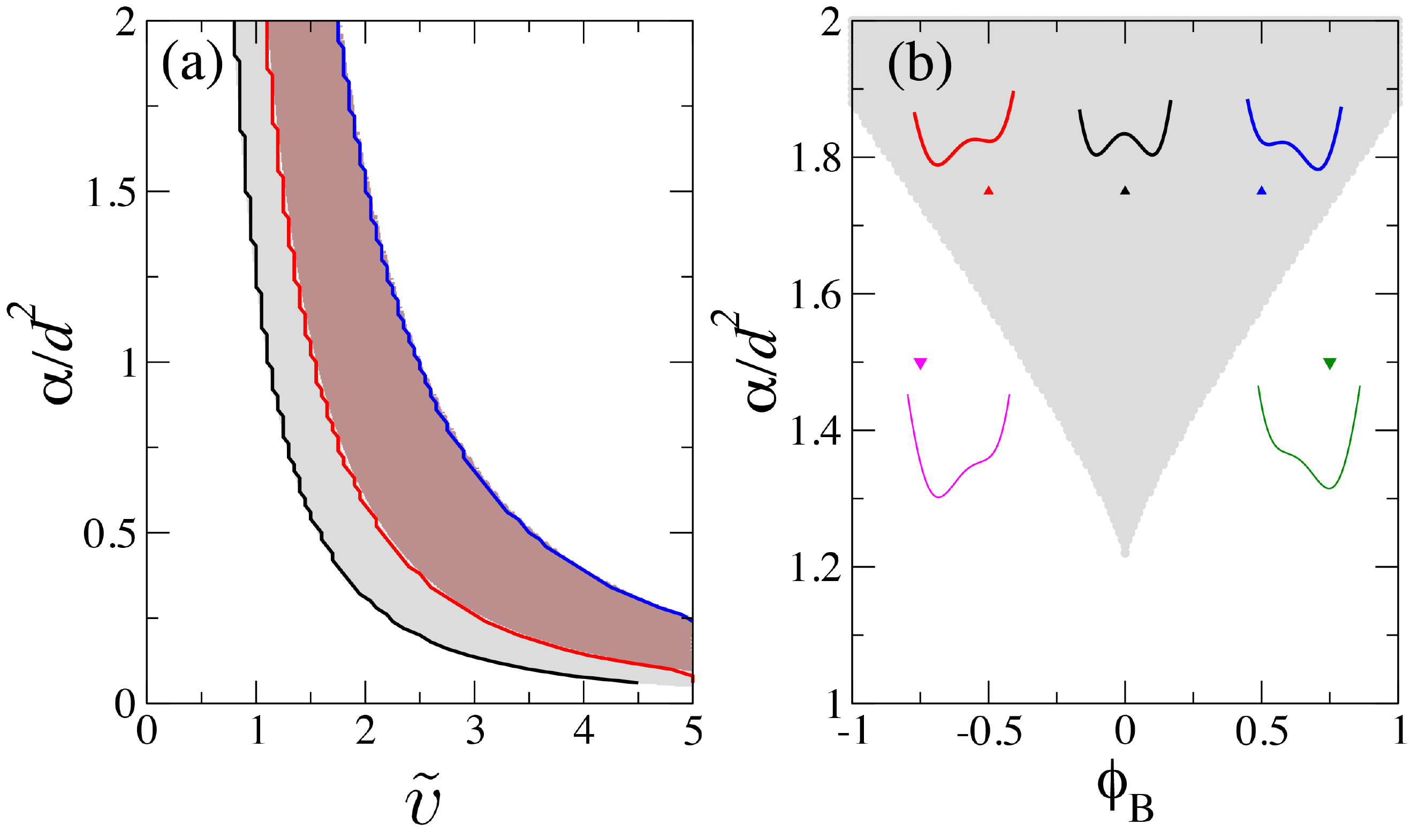}
\caption{Phase diagram for bi-stability in the parameter domain $(\tilde{v},\alpha/d^2)$ under $\phi_B=0$ for (a) and $(\phi_B,\alpha/d^2)$ with $\tilde{v}=1$ for (b) by using adiabatic approximation, Eq.(\ref{qdist}). The brown colored area in (a) represents the shuttling-promoted switching regime, $\tilde{v}_2<\tilde{v}<\tilde{v}_3$. The details for computing the boundaries of each domain are given in the Supplementary materials. Thick lines denote numerical solution of Eq.(\ref{xtil}) and Eq.({\ref{qtil}}) describing the evolution of critical voltages $\tilde{v}_1$ (black), $\tilde{v}_2$ (red), and $\tilde{v}_3$ (blue) as a function of applied bias $\tilde v$. 
The shape of electromechanical potentials in (b) corresponds to the points $\phi_B=0$ (black), $-0.5$ (red), and $0.5$ (blue) at $\alpha=1.75$, {\color{black} and $\phi_B=-0.75$ (magenta), $0.75$ (green) at $\alpha=1.5$}}
 \label{fig:f4}
 \end{figure}

In conclusion, we have reported current-switching device promoted by shuttling phenomenon based on magnetically controllable bi-stability based on strong NEM coupling.
The NEM coupling gives rise to double-well shaped electromechanical potential controlled by bias voltage between source and drain.
Based on the electromechanical pumping provided by shuttling phenomenon, the switch can transfer huge current power as an active device.
We have performed both numerical and analytical analysis and found regime of shuttling instability.

It worth noting that similar instability and transient from damping NEM oscilations (cooling regime) to self-sustained large-amplitude shuttling (heating regime) may be realized for "Kondo shuttling" \cite{kis06,prlkiselev} between metallic lead and long metallic cantilever with attached nano-island \cite{kondo}. It is tempting to unite spin and charge related switching  mechanisms in the same system, e.g. in a three-terminal device with two Kondo shuttles.

The principal scheme of device proposed in this letter may be useful not only for switching application in electronic circuits  but also for testing magnetically fine-tuned multi-stability in non-linear system as well as toy model of dynamic critical phenomena in dissipative systems.

We appreciate fruitful discussions with K.-H. Ahn, Hee Chul Park and S. Ludwig.
The research of KK was partially supported by ISF Grant No. 400/12.
The work of RIS and LYG was supported in part by Swedish VR.

\end{document}